\documentclass[letter,twocolumn]{jpsj2}

\newcommand{\Journal}[4]{#1 \textbf{#2} (#4) #3}
\newcommand{\PRev}{Phys. Rev.}
\newcommand{\JPSJ}{J. Phys. Soc. Jpn.}
\newcommand{\be}{\begin{equation}}
\newcommand{\ee}{\end{equation}}

\newcommand{\cd}{Cd$_2$Re$_2$O$_7$}

\title{Structural Order Parameter in the Pyrochlore Superconductor \cd}

\author{I. A. \textsc{Sergienko} 
\thanks{E-mail address: isergien@physics.mun.ca} and S. H. \textsc{Curnoe}}
\inst{Department of Physics and Physical Oceanography, 
Memorial University of Newfoundland, St. John's, NL, A1B 3X7, Canada}

\abst{It is shown that both structural phase transitions in \cd, which occur at 
$T_{s1}=200~K$ and $T_{s2}=120~K$, are due to an instability of the Re tetrahedral 
network with respect to the same doubly degenerate long-wavelength phonon mode. 
The primary structural order parameter transforms 
according to the irreducible representation $E_u$ of the point group $O_h$.
We argue that the transition at $T_{s1}$ may be of second order, in accordance
with experimental data. We obtain the phase diagram in the space of phenomenological
parameters and propose a thermodynamic path that \cd\ follows upon cooling.
Couplings of the itinerant electronic system and localized spin states in 
pyrochlores and spinels to atomic displacements are discussed.}

\kword{pyrochlore, \cd, structural phase transitions, order parameter}

\begin{document}

\maketitle

The pyrochlore structure with localized spins is a well known example of a
three-dimensional geometrically frustrated system~\cite{anderson}.
Pyrochlore compounds with itinerant electrons have also recently attracted much 
interest due to the discovery of the first pyrochlore superconductor 
\cd\cite{sakai}.
Type-II superconductivity appears below 
$T_c$=1.5 K\cite{sakai,hanawa,jin1,lums}.
In addition, \cd, an ideal pyrochlore at room temperature (space group 
$Fd\overline3m$) exhibits two structural phase transitions, at 
$T_{s1}$=200 K and at $T_{s2}$=120 K. 
Both transitions are accompanied by anomalous transport properties. The
electrical resistivity 
is weakly temperature dependent above 200 K, but shows a dramatic change in 
slope at $T_{s1}$, and then continues to decrease down to the onset of 
superconductivity\cite{sakai, hanawa, hiroi}. Hall coefficient measurements reveal 
that the conductivity is hole-like above $T_{s1}$ and electron-like below 
$T_{s1}$\cite{hiroi2}. Also, the magnetic susceptibility as a function of
temperature drops below $T_{s1}$~\cite{sakai,hanawa,jin,sakai1}.
Re nuclear quadrupole resonance experiments find no evidence of 
magnetic order in the low temperature phases of \cd\cite{vyas}.
The phase transition at $T_{s1}$ has been shown to be of second-order and 
specific heat measurements reveal a
pronounced $\lambda$-type anomaly\cite{hiroi1}. 
Evidence of a first-order phase transition at $T_{s2}$ has been observed as a 
small but clear temperature hysteresis of electrical resistivity\cite{hiroi1}. 
The transition is also accompanied by small anomalies in the Hall coefficient 
and thermoelectric power but there is no change in the 
susceptibility\cite{hiroi2,huo}.

The first x-ray diffraction study on \cd\ led to the conclusion that the 
space group symmetry for $T<T_{s1}$ is $F\overline43m$. 
But recent convergent-beam electron diffraction\cite{tsuda,tsuda1} and x-ray 
diffraction 
experiments\cite{yamaura} together with a Re NMR\cite{vyas} study clearly
show that both inversion and three-fold symmetry is broken in both low
temperature phases. 
Both phases are tetragonal and most probably the space group for 
$T_{s2}<T<T_{s1}$ (hereafter referred to as phase II) is $I\overline4m2$, 
while that for $T<T_{s2}$ (phase III) is $I4_122$\cite{yamaura}. 

In this Letter we present a group theoretical analysis of phonons
in pyrochlores and closely related spinels. We find that both structural 
phase transitions in \cd~originate from an instability of the pyrochlore
lattice with respect to the same long wavelength phonon mode, which
is doubly degenerate in the room temperature phase (phase I).
The I-II transition may be of second order despite the fact that 
$I\overline4m2$ is not a maximal subgroup of $Fd\overline 3m$.
We model the phase transition sequence using Landau theory.
Finally, we discuss the possibility of Jahn-Teller-like couplings of electronic 
band structure, as well as localized spin states in pyrochlores and spinels, 
to atomic displacements.
 
\begin{table}[t]
\caption{\label{phonons}Brillouin zone centered phonon modes in (a) the pyrochlore 
Cd$_2$Re$_2$O(1)$_6$O(2) and (b) a generic spinel compound AB$_2$O$_4$. 
The numbers in columns (a) and (b) are the number of atoms per fcc site.}
\begin{tabular}{lll}
\hline
\qquad (a) & (b) & \qquad Modes\\
\hline
4 Re, 4 Cd & 4 B & $A_{2u}+E_u+F_{2u}+2F_{1u}$\\
12 O(1) & & $A_{1g}+A_{2u}+E_g+E_u+2F_{1g}+3F_{2g}$\\
&& $+3F_{1u}+2F_{2u}$\\
2 O(2) & 2 A & $F_{2g}+F_{1u}$\\
& 8 O & $A_{1g}+A_{2u}+E_g+E_u+F_{1g}+2F_{2g}$\\
&& $+2F_{1u}+F_{2u}$\\
\hline
\end{tabular}
\end{table}

There is no multiplication of the pyrochlore primitive cell in the low
temperature phases of \cd\cite{yamaura}, which implies that Brillouin 
zone-centered long-wavelength phonons are crucial for the phase transitions
at $T_{s1}$ and $T_{s2}$. 
These phonons are most conveniently classified according to the irreducible 
representations of the point group $O_h$.
Table~\ref{phonons} lists the long-wavelength phonons of the atomic
displacements in pyrochlores and spinels. 
It is well established by x-ray studies that inversion symmetry is broken in 
both phases II and III\cite{hiroi, hiroi1, yamaura}. Therefore it follows that the 
structural phase transitions are due to a lattice instability
with respect to \emph{odd} phonon modes. 
Table~\ref{ops} lists all low symmetry phases that can be described 
by order parameters (OP) spanning
odd representations of $Fd\overline3m$ at the $\Gamma$ point of the Brillouin
zone. We find in the following that both phases II and III are described by 
the $E_u$ OP, and there are no restrictions in principle for both of the I-II and 
I-III phase transitions to be of 
second order. On the other hand, the II-III phase transition can be of first order 
only, since there is no group-subgroup relationship between symmetry groups of these
phases.

\begin{table}[t]
\caption{\label{ops}Possible low-symmetry phases described by odd-parity
OP's at $\Gamma$ point of Brillouin zone. Phases accessible from $Fd\overline3m$ 
by a single second-order phase transition are written in bold typeset. 
Plain typeset indicates phases, which can be accessed by a first-order phase 
transition or a sequence of phase transitions.}
\begin{tabular}{ll}
\hline
$A_{1u}$ & $\mib{F4_132}$\\
$A_{2u}$ & $\mib{F\overline43m}$\\
$E_u$ & $\mib{I\overline4m2}, \mib{I4_122}, F222$\\
$F_{1u}$ & $\mib{I4_1md}, \mib{R3m}, Ima2, Cm, Cc, P1$\\
$F_{2u}$ & $\mib{I\overline42d}, \mib{R32}, Ima2, C2, Cc, P1$\\ 
\hline
\end{tabular}
\end{table}

\begin{figure}[t]
\caption{\label{topol}The action of $Fd\overline3m$ space group symmetry 
operations on the $E_u$ order parameter space. 
Shown are 12 vectors obtained by applying all the symmetry 
operations to a generic vector $(\eta_1, \eta_2)$ (shown dashed). The numbers
in parentheses indicate symmetry operations of the space group\cite{ITC} that 
transform the initial vector to the corresponding positions. 
Every position is reached by four symmetry operations, but except for the initial
position, only one of the four symmetry operations is given.}
\includegraphics{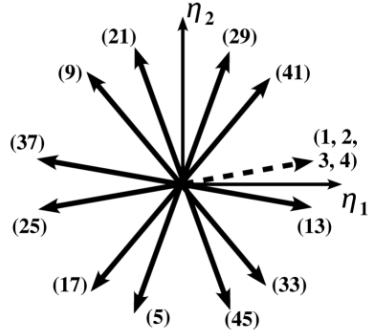}
\end{figure}

We denote the two components of the $E_u$ OP as $(\eta_1, \eta_2)$
and consider the action of the $Fd\overline3m$ space group operations 
on the space spanned by $\eta_1$ and $\eta_2$ (see Fig.~\ref{topol}). The 
topology of this space coincides with the topology of the 
plane point group $C_{6v}$.
There are two different special symmetry positions of the vector 
$(\eta_1, \eta_2)$, corresponding to the distorted phases: 
II, $(0, \eta_2)$ and III, $(\eta_1, 0)$, with space
groups $I\overline4m2$ and $I4_122$ respectively. 
Each of these two positions have 
five counterparts, which correspond to different domains. 
They are obtained by successively applying six-fold rotations in the OP space.
The generic position of $(\eta_1, \eta_2)$ corresponds to the $F222$ 
orthorhombic phase as shown in Table~\ref{ops}.

\begin{figure}[t]
\caption{\label{disps}Two adjacent tetrahedra formed by Re atoms (filled circles)
in \cd. Also shown (empty circles) are six oxygen atoms, which are the 
nearest neighbours of Re atom 1. Atoms related by an fcc translation are equilavent 
and are labelled by the same number. Arrows indicate
atomic displacements in phases (a) $I\overline4m2$ (phase II) and (b) $I4_122$
(phase III).}
\includegraphics{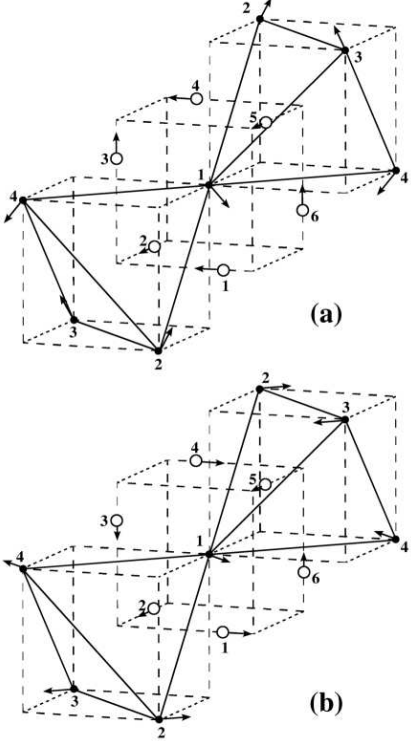}
\end{figure}

In order to reveal the nature of the structural phase transitions, we decompose
explicitly atomic displacements into irreducible representations.
Fig.~\ref{disps} shows two adjacent tetrahedra formed by Re atoms, as well as 
six oxygen atoms occuping the 48(f) Wyckoff position\cite{ITC}, which are the 
nearest neighbours of one of the Re atoms. 
%First we consider the tetrahedral network formed by the Re atoms. 
We enumerate the
four non-equivalent tetrahedral vertices, located at one fcc site as shown
on Fig.~\ref{disps}. The coordinates in the ideal pyrochlore structure are 
$(1/8,1/8,1/8)$, $(1/8,-1/8,-1/8)$, $(-1/8,1/8,-1/8)$ and $(-1/8,-1/8,1/8)$ for
%$\mathbf r^0_1=(1/8,1/8,1/8)$, $\mathbf r^0_2=(1/8,-1/8,-1/8)$,
%$\mathbf r^0_3=(-1/8,1/8,-1/8)$ and $\mathbf r^0_4=(-1/8,-1/8,1/8)$ for
Re atoms $m=1, 2, 3$ and $4$ respectively.
The components of the displacement of the $m$th Re atom from its ideal position 
%$\mathbf r^0_m$ 
are denoted $(x_m, y_m, z_m)$. Then 

\be
\label{eta_op}
\begin{split}
\eta_1&=(X-Y)/\sqrt{2},\\
\eta_2&=(X+Y-2Z)/\sqrt{6},
\end{split}
\ee
where $X=(x_1+x_2-x_3-x_4)/2$, $Y=(y_1-y_2+y_3-y_4)/2$ and 
$Z=(z_1-z_2-z_3+z_4)/2$.

To obtain the complete picture of atomic displacements in the ordered phases
we also keep track of \emph{secondary} OP's, 
which are not critical but nevertheless become
finite in the distorted phases due to coupling to the primary OP. 
In order to determine the representations corresponding to the secondary OP's,
we consider the second and third symmetric powers of the primary 
representation 
$[E_u\otimes E_u]=A_{1g}\oplus E_g$ and 
$[E_u\otimes E_u\otimes E_u]=A_{1u}\oplus A_{2u} \oplus E_u$ 
and find that $A_{2u}$ and $E_g$ phonons (listed in Table~\ref{phonons}) 
correspond to secondary OP's (the breathing mode $A_{1g}$ has been omitted). 
Of them, only $A_{2u}$ involves Re atomic displacements.
We denote the $A_{2u}$ normal coordinate as $\phi$ and obtain

\be
\label{phi_op}
\phi=(X+Y+Z)/\sqrt{3}.
\ee
This mode couples to the the primary order parameter through the invariant 
combination $(\eta_2^3-3\eta_1^2\eta_2)\phi$. Thus, in the first 
approximation, in phase II, $\phi$ is proportional to $\eta_2^3$, and in phase III 
$\phi$ is exactly zero. We have also found the normal coordinates for 
the other Re modes listed in Table~\ref{phonons}, which
vanish exactly according to the symmetry of phases II and III. 
The inverse transformation yields:

\be
\begin{split}
\label{disp_in_ops}
x_1=x_2=-x_3=-x_4&=(\sqrt3\eta_1+\eta_2+\sqrt2\phi)/\sqrt{24}\\
y_1=-y_2=y_3=-y_4&=(-\sqrt3\eta_1+\eta_2+\sqrt2\phi)/\sqrt{24}\\
-z_1=z_2=z_3=-z_4&=(2 \eta_2-\sqrt2 \phi)/\sqrt{24}.
\end{split}
\ee
It is worthwhile to note here the major difference between the 
tetrahedral network and a single tetrahedral molecule\cite{yamashita}. 
In the latter, the normal coordinate~(\ref{phi_op}) corresponds to the $A_1$ 
breathing mode and, unlike in crystals, may be omitted in most cases. 

Thus we find that in phase II, $x_1=y_1\approx -z_1/2$ and 
in phase III, $x_1=-y_1$ and $z_1=0$. The directions of the displacements are 
shown in 
Fig.~\ref{disps}. They are in complete accordance with the results of
x-ray diffraction studies\cite{yamaura}. In particular, in phase II, the two 
adjacent tetrahedra have different volumes, 
$\Delta V/V \approx \pm 8\eta_2^2$, where different signs should be taken for
neighbouring tetrahedra, and there are four different Re-Re bond lengths, 
$\Delta l_{1,2}/l\approx (\pm \sqrt{2/3} \eta_2+3\eta_2^2)$ and 
$\Delta l_{3,4}/l\approx \pm \sqrt{8/3}\eta_2$. In phase III the
volumes of the tetrahedra are the same and there are three different 
Re-Re bond lengths, 
$\Delta l_{1,2}/l\approx (\pm \sqrt2 \eta_1+\eta_1^2)$ and 
$\Delta l_3/l\approx 4\eta_1^2$.

The Cd atoms in \cd\ form a similar tetrahedral network shifted with respect to the
Re network by a $(1/2,1/2,1/2)$ translation. Thus the directions of the Cd atomic 
displacements in the distorted phases are exactly the same as those for Re atoms, 
although their absolute values differ in general. 

Similar considerations for the 12 oxygen O(1) atoms located at one fcc site 
yield

\be
\label{disp_in_ops_O}
\begin{split}
x'_1=-x'_{10}&=(\sqrt3\eta'_1-\eta'_2+\sqrt2\phi'+\xi'_1+\sqrt3\xi'_2)/
\sqrt{24}\\
x'_4=-x'_7&=(\sqrt3\eta'_1-\eta'_2+\sqrt2\phi'-\xi'_1-\sqrt3\xi'_2)/
\sqrt{24}\\
y'_2=-y'_{11}&=(-\sqrt3\eta'_1-\eta'_2+\sqrt2\phi'+\xi'_1-\sqrt3\xi'_2)/
\sqrt{24}\\
y'_5=-y'_8&=(-\sqrt3\eta'_1-\eta'_2+\sqrt2\phi'-\xi'_1+\sqrt3\xi'_2)/
\sqrt{24}\\
z'_3=-z'_{12}&=(2 \eta'_2+\sqrt2 \phi'-2\xi'_1)/\sqrt{24}\\
z'_6=-z'_9&=(2 \eta'_2+\sqrt2 \phi' +2\xi'_1)/\sqrt{24},
\end{split}
\ee
while all of the other components of the displacements vanish. 
Fig.~\ref{disps} shows six of the twelve oxygen atoms with their displacements in
phases II and III. The undistorted coordinates of atom 1 are $(v,0,0)$, where the
oxygen structural parameter is $v\approx 0.192$\cite{hiroi2}. 
The coordinates of the $n$th atom $n>6$ are
obtained by adding a $(1/4,1/4,1/4)$ translation to the coordinates of atom 
$n-6$. By analogy with the normal coordinates of the Re atoms, we denote the
$E_u$ and $A_{2u}$ modes by $(\eta'_1, \eta'_2)$ and $\phi'$ respectively. We also
introduce normal coordinates for $E_g$ $(\xi'_1, \xi'_2)$ since this mode
is present in O(1) displacements (see Table~\ref{phonons}), which is 
a secondary OP, coupled to the primary OP as 
$(\eta_1^2-\eta_2^2)\xi'_1-2\eta_1\eta_2\xi'_2$.
It follows that $\xi'_2=0$ and $\xi'_1\ne0$ in both phases II and III.

A Landau-type model with $C_{6v}$ topology was first studied by 
Lifshitz\cite{lift}, who considered a sixth-order expansion of the 
thermodynamic potential. Details of calculations and general formalism 
can be found in\cite{gufan, toled}. 
Here we briefly summarise the results and apply them to \cd. 
Only even-order polynomial invariants can be constructed from the OP 
components $\eta_1 \text{ and } \eta_2$. There is only one isotropic fourth-order 
term $(\eta_1^2+\eta_2^2)^2$, which does not lift degeneracy of ordered 
states. The first anisotropic term  appears in sixth order,
$(\eta_1^3-3\eta_1\eta_2^2)^2$, but as is easily
seen from the following, the sixth-order model still 
fails to lift degeneracy between ordered phases along the line 
of the II--III phase transition. Hence we consider the eighth-order Landau 
expansion

\be
\label{pot}
\begin{split}
F=&a_1(\eta_1^2+\eta_2^2)+a_2(\eta_1^2+\eta_2^2)^2+a_3(\eta_1^2+\eta_2^2)^3\\
&+b_1(\eta_1^3-3\eta_1\eta_2^2)^2+a_4(\eta_1^2+\eta_2^2)^4\\
&+c(\eta_1^2+\eta_2^2)(\eta_1^3-3\eta_1\eta_2^2)^2.
\end{split}
\ee

\begin{figure}[t]
\caption{\label{pdc}(a), (b) Phase diagrams of potential~(\ref{pot}) for
$a_2=a_3=a_4=1$. (a) $c=0.5>0$, (b) $c=-0.5<0$. Solid and dashed lines are lines of first 
and second order phase tarnsitions respectively. Dot-dasehd lines are metastability 
boundaries of phases II and III.
(c), (d) Contour plots of 
$F(\eta_1,\eta_2)$ for the case (b) and $a_1=-2$, (c) $b_1=0.5$ (phase II),
(d) $b_1=-0.2$ (phase III).}
\includegraphics{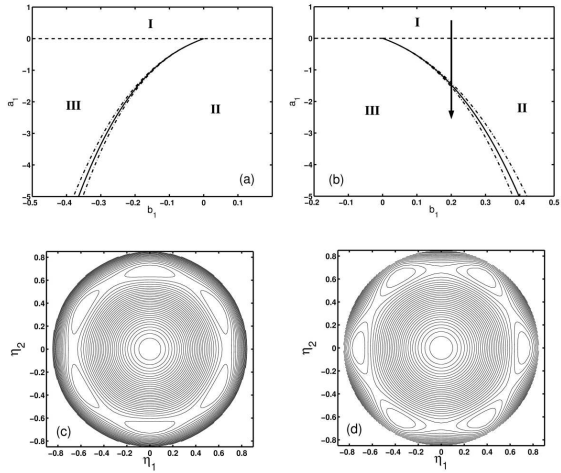}
\end{figure}

Using standard minimization procedure, one finds that when $a_1=0$ a
second-order phase transition from phase I occurs to either phase II or III, 
depending on the sign of $b_1$. In Fig.~\ref{pdc} we plot phase diagrams
in the $a_1b_1$ plane, obtained for $a_2>0$, $a_3>0$, $a_4>0$ and different signs
of $c$. Overall stability of the potential~(\ref{pot}) requires that $a_4+c>0$. 
Contour plots of the potential as a function of $\eta_1$ and $\eta_2$ in 
phases II and III are shown in Fig.~\ref{pdc}(c) and~\ref{pdc}(d). 
One can clearly see for both phases six minima corresponding to six tetragonal 
domains.

The potential minima for phase II exist up to the line
\be\label{lII}
a_1=2a_2b_1/c-3a_3(b_1/c)^2+4a_4(b_1/c)^3,
\ee
and for phase III
\be\label{lIII}
a_1=2a_2b_1/c-3a_3(b_1/c)^2+(4a_4+c)(b_1/c)^3.
\ee
Equating potentials of phases II and III, we obtain the line of the first-order
II--III phase transition,
\be
a_1=2a_2b_1/c-3a_3(b_1/c)^2+(4a_4+c/2)(b_1/c)^3.
\ee

Fitting data about the I--II phase transition line from high pressure resistivity 
experiments\cite{hiroi} to $a_1=0$, we
obtain $a_1=\alpha[T-200\text{ K}+P\cdot 40\text{ K/GPa}]$, 
where $P$ stands for pressure and $\alpha$ is an undetermined parameter. 
There is not enough data to determine unambiguously where the 
II--III transition line is located in the $P$-$T$ phase diagram.
However, we can speculate that the system follows, on cooling, the path indicated
by the arrow in Fig.~\ref{pdc}(b).
It follows from Eqs.~(\ref{lII}) and~(\ref{lIII}) that the width of the area 
where two phases II and III coexist is of third order of 
magnitude in distance from the triple point $a_1=b_1=0$,
which is consistent with the smallness of temperature hysteresis observed
in resistivity mesaurements\cite{hiroi1}.

In an applied magnetic field $\mib{H}$ additional terms appear in the 
thermodynamic potential, resulting from magnetization energy and coupling
between the magnetization $\mib{M}$ and the structural OP

\be
\Delta F_M = \gamma (\eta_1^2+\eta_2^2)\mib{M}^2 + A\mib{M}^2-\mib{MH},
\ee
where we neglect higher-order terms of magnetic anysotropy.
Then the magnetic susceptibility is
\be
\chi^{-1}=\chi_0^{-1}+2\gamma(\eta_1^2+\eta_2^2),
\ee
where $\chi_0=(2A)^{-1}$ is the susceptibility of the high symmetry phase. 
Assuming $\gamma>0$, this expression accounts for the susceptibility decrease 
observed in phases II and III.

Thus we have obtained the full description of atomic displacements occuring in \cd\ 
at the structural phase transitions at $T_{s1}$ and $T_{s2}$. 
The phenomenological treatment presented in this Letter
is in complete agreement with experimental data available to date. 
On the other hand, a detailed account of the intriguing transport 
properties of this compound requires a consideration of the underlying microscopic
physics. Dramatic changes in resistivity, thermoelectric power, Hall 
coefficient and magnetic susceptibility at the structural phase 
transitions, point to the importance of coupling of itinerant electrons to 
long-wavelength phonons. This fact sets \cd\ apart from other pyrochlores. 
For example, the closely
related compound Cd$_2$Os$_2$O$_7$ demostrates a metal-insulator transition 
near 225 K, accompanied by no structural changes\cite{sleight, mandrus}.
As we have shown, the displacements of all Cd, Re and O(1) atoms include normal 
modes, which have the symmetry of the structural OP. Recent band structure 
calculations for the ideal pyrochlore phase\cite{harima,singh} show that the 
transport properties of \cd\ are defined mostly by Re 5d electrons. 
Therefore we conclude that Re displacements play major role in the structural 
transitions and that the transitions occur due to instability of the Re tetrahedral 
sublattice with respect to the phonon mode~(\ref{eta_op}). 
In addition, a gain in itinerant electronic energy due to a Jahn-Teller effect
may favour the phase transformations. 
In particular, spin-orbit coupling has been shown to be important for band
dispersion near the Fermi level\cite{harima,singh}. This coupling results
in a complete lifting of spin degeneracy of electronic levels at generic points
of the Brillouin zone below $T_{s1}$ due to loss of inversion symmetry\cite{jones}.

We would also like to comment on the coupling of structural order parameters 
with spins in localized spin systems in pyrochlores 
and spinels. Yamashita and Ueda\cite{yamashita} considered a single tetrahedron 
of spin-1 atoms and found that the degeneracy of spin 
states can be lifted by a Jahn-Teller mechanism. 
It was possible due to presence in the Hamiltonian terms 
of type $(\partial J/\partial Q_\alpha) Q_\alpha~(s_is_j)_\alpha$, where $J$
is the direct exchange coupling between two tetrahedron vertices, 
$Q_\alpha$ is a linear combinations of atomic 
displacements, corresponding to normal modes, and $(s_is_j)_\alpha$ stands for a
symmetry adopted bilinear combination of spin-1/2 operators located at $i$ and
$j$ vertices. Our analysis of
phonon modes in the lattice of corner-shared tetrahedra shows that sum of two
such terms for two adjacent tetrahera vanishes
since every vertex is a center of inversion $I$. This is because vertex located 
spins transform to equivalent spins under inversion: $Is_i=s_i, i=1,\ldots,4$ and 
thus their combinations are of even parity, while phonon modes of vertex atoms are
odd under inversion.

On the other hand, there are still three possibilities of frustration removing
due to spin-displacement coupling:
(i) Unlike tetrahedra-corner atoms, oxygen O(1) displacements contain even
phonon modes of $E_g$ and $F_{1g}$ symmetry ($A_{1g}$ is trivial breathing 
mode). We denote the normal coordinate corresponding to one of these modes by $Q_1$.
Then the following coupling is possible:  
$(\partial J_1/\partial Q_1) Q_1(s_is_j)$, where $J_1$ is the superexchange 
coupling via nearest-neighbouring oxygen atoms.
(ii) The structural displacements or the spins are spatially modulated (non-zero
wave vector), so that translations of the fcc lattice are not preserved.
(iii) Spin states couple to quadratic a form of the tetrahedral displacements and 
play the role of secondary order parameters  
$(\partial^2 J/\partial Q_m \partial Q_n) Q_mQ_n(s_is_j)$. In this case, 
spin ordering may be driven by structural distortions but the mechanism 
cannot be classified as Jahn-Teller.

In summary, we have shown that the structural phase transitions in \cd\ 
at 200 K and 120 K can both be described by a single, two-component OP, which is the
displacements of Re atoms transforming according to the representation $E_u$ of
the point group $O_h$. We have also considered all secondary OP's, which can couple
to the primary OP. A Landau model with terms up to eighth order in the 
displacements is proposed in order to account for both phase transitions and the
resulting phenomenology is in good agreement with experiments.

%\newpage

\end{document}